\documentclass[
reprint,           
superscriptaddress,
amsmath,           
amssymb,           
aps,               
prd,               
notitlepage,       
longbibliography,  
floatfix,          
nofootinbib,
]{revtex4-1}

\usepackage{tensor}     
\usepackage{graphicx}   
\usepackage[
colorlinks=true,        
citecolor=blue,         
linkcolor=blue,         
urlcolor=blue           
]{hyperref}             
\usepackage{bm}         
\usepackage{xcolor}     
\usepackage{array}      
\usepackage{lipsum}
\usepackage{color}      
\usepackage[utf8]{inputenc} 
\usepackage[section]{placeins} 
\usepackage{multirow}

\newcommand{\nc}{\newcommand*} 

\nc{\al}{\alpha}
\nc{\s}{\sigma}
\nc{\kp}{\kappa}
\nc{\dt}{\delta}
\nc{\Dt}{\Delta}
\nc{\Ld}{\Lambda}
\nc{\p}{\partial}
\nc{\Gm}{\Gamma}
\nc{\om}{\omega}
\nc{\Om}{\Omega}
\nc{\rd}{\mathrm{d}}
\nc{\Od}{\mathcal{O}} 
\def\({\left(}
\def\){\right)}
\def\[{\left[}
\def\]{\right]}
\def\e{\begin{equation}}
\def\q{\end{equation}}
\def\m{\begin{eqnarray}}
\def\n{\end{eqnarray}}
\nc{\Eq}[1]{Eq.~\eqref{#1}}     
\nc{\Fig}[1]{Fig.~\ref{#1}}     
\nc{\Table}[1]{Table~\ref{#1}}  
\nc{\Sec}[1]{Sec.~\ref{#1}}     
\nc{\Msun}{M_\odot}             
\nc{\fpbh}{f_{\mathrm{PBH}}}    
\nc{\fpbhn}{f_{\mathrm{pbh0}}}    
\nc{\mR}{\mathcal{R}} 
\nc{\seq}{\sigma_{\mathrm{eq}}}
\nc{\ogw}{\Omega_{\mathrm{GW}}}
\nc{\gpcyr}{\mathrm{Gpc}^{-3}\,\mathrm{yr}^{-1}}
\nc{\lvc}{LIGO-Virgo} 
\nc{\SNR}{\mathrm{SNR}} 
\nc{\mmin}{{m_{\mathrm{min}}}}
\nc{\mmax}{{m_{\mathrm{max}}}}
\nc{\Mmin}{{M_{\mathrm{min}}}}
\nc{\fmin}{{f_{\mathrm{min}}}}
\nc{\VT}{\mathrm{VT}}
\nc{\rhoGW}{\rho_{\mathrm{GW}}}
\nc{\vth}{\vec{\theta}}
\nc{\vd}{\vec{d}}
\nc{\vla}{\vec{\lambda}}
\nc{\Nobs}{N_{\mathrm{obs}}}
\nc{\av}[1]{\langle #1 \rangle} 
\nc{\km}{\mathrm{km}}
\nc{\Mpc}{\mathrm{Mpc}}
\nc{\Tobs}{T_{\mathrm{obs}}}
\nc{\Ntemp}{N_{\mathrm{temp}}}
\nc{\fyr}{f_{\mathrm{yr}}}
\nc{\addref}{[\textcolor{red}{add ref}] } 
\nc{\eg}{\textit{e.g.~}}
\nc{\app}{\approx}
\nc{\hf}{\frac{1}{2}}
\nc{\discuss}{\textcolor{red}{Add discussion here!}}
\nc{\red}[1]{\textcolor{red}{#1}}

\begin{document}
	
\title{The Shadow of Supertranslated Schwarzschild Black Hole}
	
\author{Qing-Hua Zhu}
\email{zhuqh@itp.ac.cn} 
\affiliation{CAS Key Laboratory of Theoretical Physics, 
		Institute of Theoretical Physics, Chinese Academy of Sciences,
		Beijing 100190, China}
\affiliation{School of Physical Sciences, 
		University of Chinese Academy of Sciences, 
		No. 19A Yuquan Road, Beijing 100049, China}
	
\author{Yu-Xuan Han}
\email{hanyuxuan@itp.ac.cn} 
\affiliation{CAS Key Laboratory of Theoretical Physics, 
		Institute of Theoretical Physics, Chinese Academy of Sciences,
		Beijing 100190, China}
\affiliation{School of Physical Sciences, 
		University of Chinese Academy of Sciences, 
		No. 19A Yuquan Road, Beijing 100049, China}	
\author{Qing-Guo Huang}
\email{Corresponding author: huangqg@itp.ac.cn}
\affiliation{CAS Key Laboratory of Theoretical Physics, 
		Institute of Theoretical Physics, Chinese Academy of Sciences,
		Beijing 100190, China}
\affiliation{School of Physical Sciences, 
		University of Chinese Academy of Sciences, 
		No. 19A Yuquan Road, Beijing 100049, China}
\affiliation{School of Fundamental Physics and Mathematical Sciences
		Hangzhou Institute for Advanced Study, UCAS, Hangzhou 310024, China}

\date{\today}
	
\begin{abstract}
	The supertranslated black hole proposed by Hawking, Perry, and Strominger might provide a resolution to the information paradox, which is usually defined by a complicated  space-time metric with even less space-time symmetries compared to Kerr black hole. In this paper, we figure out the shadow for the supertranslated Schwarzschild black hole by making use of supertranslated 4-velocities and the trajectories of the light rays. Based on this approach, we find that the photon sphere gets distorted due to the supertranslation hairs and the position of the shadow on the projection plane is shifted by the supertranslation vector, but the size and shape of the shadow remain the same as those of Schwarzschild black hole.   
\end{abstract}
	
\maketitle
\newpage
 
\section{introduction}
In 1975, Hawking pointed out that black holes will end theirs lives with evaporation \cite{Hawking:1975vcx}, which significantly supported the picture of the Bekenstein entropy, that a black hole can be reckon as a thermodynamic system \cite{Bekenstein:1972tm}.  However, a question soon arose, due to its breaking the law of information conservation in conflict with the quantum theory, which is the so-called black hole information paradox. About a half century passed, in 2016, Hawking, Perry, and Strominger  (HPS) revisited this problem, and proposed soft hairy black hole that might provide a resolution to the information paradox \cite{Hawking:2016msc,Hawking:2016sgy}.


The studies on the soft hairy black hole would involve the global spacetime structure from the event horizon to the asymptotically flat zone. Therefore, it is interesting to examine whether the influence from the  soft hair can be observed. Recently, it becomes more promising, because
Event Horizon Telescope (EHT) presented the images of the black holes in M87 \cite{EventHorizonTelescope:2019dse,EventHorizonTelescope:2019ggy,EventHorizonTelescope:2019jan,EventHorizonTelescope:2019pgp,EventHorizonTelescope:2019ths,EventHorizonTelescope:2019uob}, and just recently in our galaxy \cite{EventHorizonTelescope:2022xnr}, and opened a new way to investigate the physics in strong field regime of gravity for the black holes. 

One can interpret the images of the black holes in this way. A black hole can not only bend lights but also swallow them. As a result, a shaded zone exists in the field of vision of an observer, and is thus called the black hole shadow. Theoretically, it was developed based on the works of Synge \cite{Synge:1966okc} and Bardeen \cite{Bardeen:1973tla}. By extracting information of a black hole from its shadow, one can tell the difference of the spacetime geometry for different black holes, and provide a way to explore properties of the black hole, such as reflection coefficients on the horizon \cite{Rummel:2019ads}, quantum structure \cite{Giddings:2016btb,Giddings:2019jwy} and naked singularity \cite{Bambhaniya:2021ybs}, or test general relativity \cite{Berti:2015itd,Psaltis:2018xkc,Glampedakis:2021oie} and the No-hair theorem \cite{Broderick:2013rlq,Psaltis:2015uza,Lin:2022ksb}.

Due to the interests of information paradox, the shadow of the soft hairy black hole has been  explored in  \cite{Lin:2022ksb,Sarkar:2021djs}. Since the soft hairy black hole has less space-time symmetries compared to the parameterized Kerr black hole \cite{Johannsen:2013vgc,Johannsen:2015hib}, the authors  simulated the light rays for  calculating the shadow  \cite{Lin:2022ksb} in which the parameterized metric is limited to some specific types of soft hairy black holes. The supertranslation hair, as a type of soft hair, is of great interesting, because the supertranslated black hole carries conservation charges at classical level \cite{Hawking:2016sgy}, and thus might have observable effect in classical physics. Ref.~\cite{Sarkar:2021djs} investigated the photon sphere of the supertranslated Schwarzschild black hole in the equatorial plane, which suggested that the changes of the photon sphere might lead to a distinguishable shadow. 

In this paper, we will investigate the shadow of supertranslated Schwarzschild black hole that is proposed by Hawking, Perry and Strominger \cite{Hawking:2016msc}.
Different from pioneers' works in \cite{Lin:2022ksb,Sarkar:2021djs}, we do not limit to the axisymmetric cases. We calculate the shadow of the supertranslated Schwarzschild black hole by making use of supertranslated 4-velocities and the trajectories of the light rays, and find that the photon sphere get distorted due to the supertranslation hairs. The position of the shadow on the projection plane is shifted by the supertranslation vectors, while the size and shape of the shadow remain the same as those of Schwarzschild black hole. 
 
The rest of the paper is organized as follows. In Section~\ref{II}, we list results of supertranslated Schwarzschild black hole proposed by HPS. In Section~\ref{III}, we calculate the 4-velocities and trajectories of the bending light rays around the supertranslated Schwarzschild black hole. In Section~\ref{IV}, we show the shadow of the supertranslated Schwarzschild black hole. In Section~\ref{V}, the conclusions and discussions are summarized.

\section{Schwarzschild black hole with supertranslation hair}\label{II}

As proposed by HPS \cite{Hawking:2016sgy}, the supertranslation hair on Schwarzschild black hole is given by introducing the supertranslations, 
\e
g_{\mu \nu}^{{ (\rm hairy)}}  =  g_{\mu \nu}^{{(\rm  bald)}}
+\mathcal{L}_{\zeta} g_{\mu \nu}^{{(\rm  bald)}}~, \label{1}
\q
which is also the procedure of implanting supertranslation hair on Schwarzschild black hole. The quantities in Eq.~(\ref{1}) will be illustrated in the following part of this paper. The metric of bald Schwarzschild black hole $g_{\mu \nu}^{(\rm  bald)}$ in advanced Bondi coordinate ($v$,$r$,$\theta$,$\phi$)  is given by 
\e
{\rm d} s^2  =  - \left( 1 - \frac{2 M}{r} \right)
{\rm d} v^2 + 2 {\rm d} v {\rm d} r + r^2 \gamma_{A   B} {\rm d} \Theta^A {\rm d} \Theta^B,
\q
where $M$ is the mass of the black hole, $\Theta^A$ denotes the angular coordinates $\theta$ or $\phi$.
The $\mathcal{L}_\zeta$ is the Lie derivative along the supertranslation vector $\zeta$ that is formulated as \cite{Sachs:1962zza,Barnich:2009se,Hawking:2016sgy}, 
\e
\zeta  =  f (\theta, \phi) \partial_{v} - \frac{1}{2} D^2 f
(\theta, \phi) \partial_r + \frac{1}{r} D^A f (\theta, \phi) \partial_A~. \label{2}
\q
where $D$ is the covariant derivative with respect to the
metric on the unit two sphere, and the $f(\theta,\phi)$ is  an arbitrary function that is proportional to the first order weak-field expansion. Here, the $\zeta$ is also the asymptotical Killing vector in Schwarzschild space-time. Expanding the $f(\theta,\phi)$ into spherical harmonics,
\e
f(\theta,\phi)=\sum_{lm}a_{lm}Y_{lm}(\theta,\phi)~, \label{4}
\q
one can find infinite number of choices of $l$ and $m$ that can give different asymptotical Killing vectors.
In particular, in the case of $l=0$, $\zeta$ is generator of the temporal translation. And in the case of $l=1$ and $m=-1,0,1$, $\zeta$ are generators of the three spatial translations, respectively. Finally, the supertranslated Schwarzschild metric $g_{\mu \nu}^{ (\rm hairy)}$ on the left hand side of Eq.~(\ref{1}) in advanced Bondi coordinate takes the form of
\m\label{5}
	{\rm d} s^2 & = & - \left( 1 - \frac{2 M}{r} - \frac{M}{r^2} D^2 f \right)
	{\rm d} v^2 + 2 {\rm d} v {\rm d} r\notag\\
	&& - D_A \left( 2 f - \frac{4 M}{r} f + D^2 f
	\right) {\rm d} v {\rm d} \Theta^A\\
	&& + (r^2 \gamma_{A   B} + 2 r   D_A D_B f - r \gamma_{A
		B} D^2 f) {\rm d} \Theta^A {\rm d} \Theta^B\notag~, 
\n
in which the metric depends on the angular coordinates $\theta$ and $\phi$ due to the function $f(\theta,\phi)$. It seems to bring complexity in the calculation of the propagation of light.

\smallskip
\section{Propagation of Light and photon sphere \label{III}}


The calculation about the shadow of a black hole based on the solutions of null geodesic equations.
From the metric shown in Eq.~(\ref{5}), it seems to be a tough task to deal with the geodesic equations. In this section, we will show that the 4-velocities and the trajectories of light rays can be obtained by the procedure of implanting the supertranslation hair. Directly solving the geodesic equations for the supertranslated Schwarzschild black hole is not necessary.

First of all, we should work out the 4-velocities of light rays for the bald Schwarzschild black hole in the Bondi coordinate. Based on the Hamilton-Jacobi method for geodesic equations \cite{Carter:1968rr}, one can obtain 4-velocities of light rays in static coordinates $(t,r,\theta,\phi)$ for the bald Schwarzschild black hole (see Eqs.~(10)--(13) in ref.~\cite{Chang:2020lmg}).
And then we can transform the 4-velocities in static coordinates into those in Bondi coordinate by using $v=t+r+2M\log(r/2M -1)$, namely 
\begin{subequations}
\m
&&p^{v, (\rm bald)}=E\left(\frac{r}{r-2M}\right)\left(1+\sqrt{\frac{r^{3}+2M\kappa-r\kappa}{r^3}}\right)~,\\
&&p^{r, (\rm bald)}=E\sqrt{\frac{r^{3}+2M\kappa-r\kappa}{r^3}}~,\\
&&p^{\theta, (\rm bald)}=\frac{E}{r^2}\sqrt{\kappa-\frac{\lambda^2}{\sin^2\theta}}~,\\ 
&&p^{\phi, {(\rm bald)}}=\frac{E\lambda}{r^2\sin^2\theta} ~,
\n 	\label{7}
\end{subequations}
where $E$, $\kappa\equiv K/E^2$ and $\lambda\equiv L/E$ are three integral constants corresponding to the energy, the total angular momentum, and  the angular momentum with respect coordinate $\phi$ for a light ray at infinity.

According to the above arguments, the 4-velocities of light rays around the supertranslated Schwarzschild black hole can be obtained by using the supertranslations,
\e
p^{\mu, (\rm hairy)} = p^{\mu, (\rm bald)} + \mathcal{L}_\zeta p^{\mu, (\rm bald)}~, 
\q
where the $\zeta$ is the supertranslation vector given in Eq.~(\ref{2}), and the components of the 4-velocities $p^{\mu, (\rm hairy)}$ are presented as follows,
\begin{subequations}
\begin{widetext}
\m
p^{v, (\rm hairy)} &=& \frac{E}{2 r^3}\Biggl( \frac{1}{\left(r - 2 M\right)^2  \sqrt{\frac{2 \kappa M + r^3 - \kappa r}{r^3}}}\Bigg(\left( - 2 \kappa M^2 + 2 Mr^3  \left(\sqrt{\frac{2 \kappa M + r^3- \kappa r}{r^3}} +1\right) +3 \kappa Mr -\kappa r^2\right)~,\notag\\
&&\left(\partial^{2}_{\theta}f + \cot \theta \partial_{\theta}f + \csc^2 \theta \partial^{2}_{\phi}f\right)\Bigg) -2r\partial_{\theta}f \sqrt{\kappa - \lambda^2 \csc^2 \theta} - 2\lambda r \csc^2 \theta \partial_{\phi}f + \frac{2 r^4 \left(\sqrt{\frac{2 \kappa M + r^3 - \kappa r}{r^3}} + 1 \right)}{r - 2 M}\Biggl)~,\\
p^{r, (\rm hairy)} &=& \frac{1}{r^2}\left( 2 \sqrt{\kappa - \lambda^2 \csc^2 \theta}\left(- \partial^{3}_{\theta}f - \cot \theta \partial^{2}_{\theta}f +\csc^2 \theta \partial_{\theta}f - \csc^2 \theta \partial_{\theta}\partial^{2}_{\phi}f +2 \cot \theta \csc^2 \theta \partial^{2}_{\phi}f\right)\right) \notag\\&& + \frac{2 \lambda   \csc^2 \theta }{r^2}\left(\partial^{2}_{\theta}\partial_{\phi}f+ \cot \theta \partial_{\theta}\partial_{\phi}f +\csc^2 \theta \partial^{3}_{\phi}f\right) +4\sqrt{\frac{2 \kappa M +r^3 -\kappa r}{r^3}}~,\\
p^{{\theta}, (\rm hairy)}  &=&  \frac{E}{r^3}\Bigg( - r   \partial^{2}_{\theta}f \sqrt{\kappa - \lambda^2 \csc^2 \theta} + \frac{\lambda^2 r \cot\theta \csc^2 \theta \partial_{\theta}f}{\sqrt{\kappa - \lambda^2\csc^2 \theta}} - \lambda r \csc^2 \theta \partial_{\theta}\partial_{\phi}f\notag\\&& +\sqrt{\kappa - \lambda^2 \csc^2 \theta}  \left(\partial^{2}_{\theta}f + \cot\theta \partial_{\theta}f + \csc^2 \theta \partial^{2}_{\phi}f\right) + r \sqrt{\kappa - \lambda^2 \csc^2\theta}\Bigg)~,\\
p^{{\phi}, (\rm hairy)}  &=&  \frac{E \csc^2 \theta}{r^3} \Bigg(-r \sqrt{\kappa - \lambda^2 \csc^2 \theta} \partial_{\theta}\partial_{\phi}f+ 2 r   \cot \theta \sqrt{\kappa - \lambda^2 \csc^2 \theta} \partial_{\phi}f  \notag\\&& - 2\lambda r   \cot \theta \partial_{\theta} f + \lambda \left(1 -r\right)   \csc^2 \theta \partial^{2}_{\phi} f + \lambda   \partial^{2}_{\theta}f + \lambda   \cot \theta \partial_{\theta} f + \lambda r \Bigg)~.
\n
\end{widetext}\label{8}
\end{subequations}
Based on the diffeomorphisms, it is not difficult to find that the 4-velocities of the light rays in Eqs.~(\ref{8}) are exactly the solutions of geodesic equations with respect to the metric in Eq.~(\ref{5}).

From the 4-velocities given in Eqs.~(\ref{8}), the worldlines of the light rays $\gamma^\mu(\lambda)$ in supertranslated Schwarzschild spacetime can be easily obtained. The geodesic equations for the worldlines of light rays $\bar{\gamma}^\mu$ in bald Schwarzschild space-time are given by
\m
\frac{\rd \bar{\gamma}^{\mu}}{\rd \lambda}&=&p^{\mu,(\rm bald)}~. \label{13-1}
\n
After implanting the supertranslation hair, the geodesic equations take the form of
\m
\frac{\rd }{\rd \lambda}(\bar{\gamma}^{\mu} + \delta{\gamma}^{\mu})&=&p^{\mu, (\rm bald)}+\mathcal{L}_{\zeta}p^{\mu, (\rm bald)}~, \label{14-1}
\n
where $\delta\gamma^\mu$ indicates a small change for the $\bar{\gamma}$ ascribed from the supertranslation hair.
Using Eqs.~(\ref{13-1}) and (\ref{14-1}), we obtain the solution $\delta\gamma^{\mu}=\zeta^{\mu}$. Therefore the worldline of a light ray around the supertranslated Schwarzschild black hole is formulated as
\m
\gamma^{\mu}=\bar{\gamma}^{\mu}+\zeta^{\mu}~. \label{15}
\n 
It shows that each event on the null geodesic is shifted  by the supertranslation vector. Since a finite $\zeta^\mu$ can not shift the ``endpoint" of a null geodesic from the singularity inside the black hole to spatial infinity, the fact that the light rays finally escape away from or fall into the black hole will not be changed after implanting supertranslation hair. 

The photon sphere for Schwarzschild black hole is the assemble of the unstable circle orbits of the light rays, which determines whether a bending light ray approaching the black hole can escape to infinity again. In this sense, it describes critical escape orbits for a light ray. Since implanting supertranslation hair does not change whether a light ray can escape to infinity or not, the 4-velocities of the light rays in critical escape orbits for the supertranslated Schwarzschild black hole can be formulated as
\e
p^{\mu, (\rm hairy)}_c = p^{\mu, (\rm bald)}|_{\kappa=27M^2} + \mathcal{L}_\zeta (p^{\mu, (\rm bald)}|_{\kappa=27M^2})~, \label{12-1}
\q
where the subscript `$c$' denotes the light rays in the critical escape orbits.

\begin{figure*}[h]
	\includegraphics[scale=0.48]{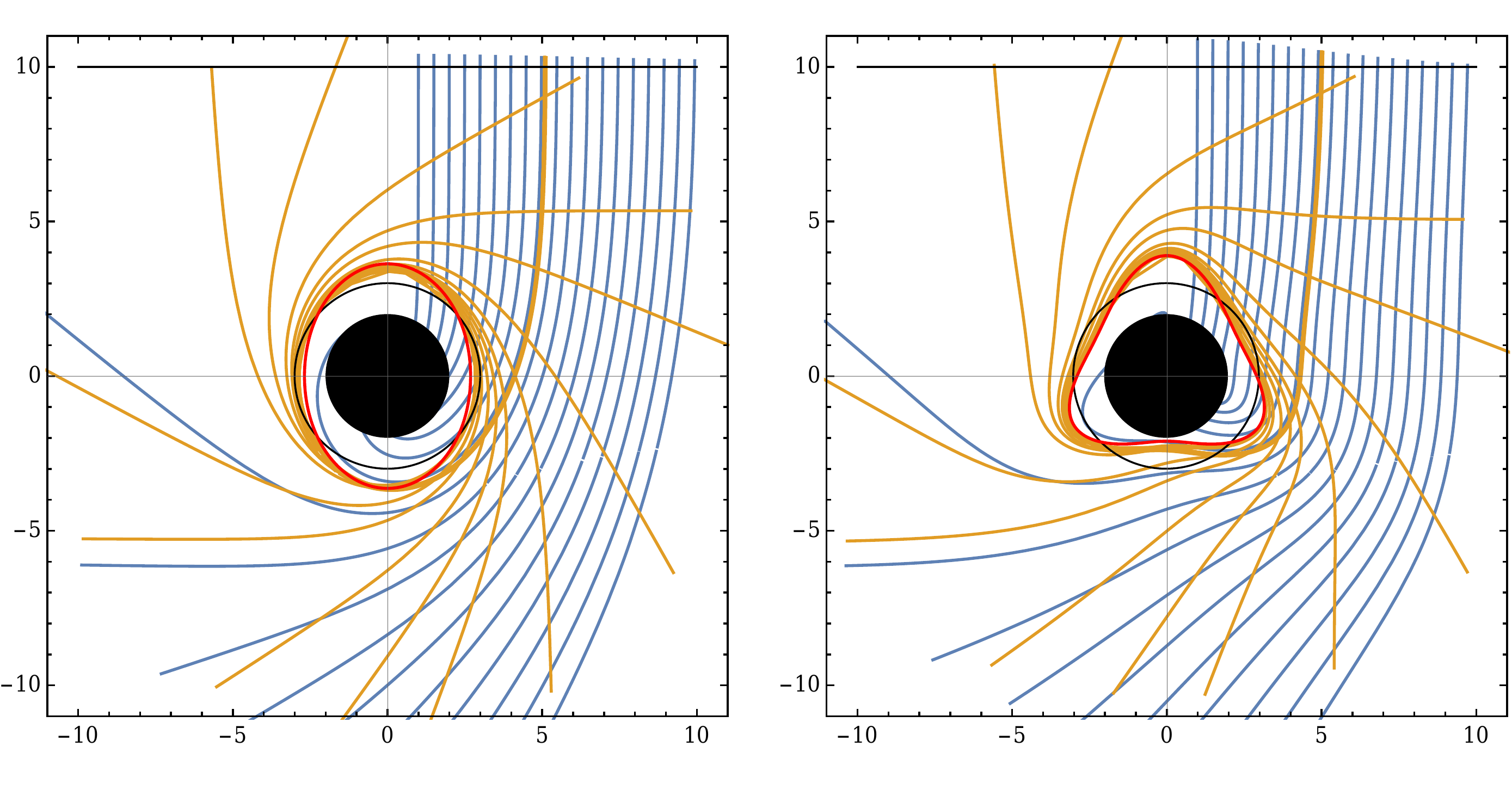}
	\caption{Ray-tracing simulations for the light rays (blue and orange curves) around a supertranslated Schwarzschild black hole with $f(\theta,\phi)=\frac{1}{5}Y_{20}(\theta,\phi)$ (left panel), and $f(\theta,\phi)=\frac{1}{5}Y_{30}(\theta,\phi)$ (right panel), respectively. The black circle represents the photon sphere for the bald black hole, and the red circle represents the photon sphere for the supertranslated Schwarzschild black hole. The orange curves represent the light rays that are almost tangent to the photon sphere, which determine the edge of the shadow on observers' projection plane. \label{F1}}
\end{figure*}

We will further illustrate Eq.~(\ref{12-1}) with ray-trace simulations.
In Fig.~\ref{F1}, we show the trajectories of the light rays $\gamma^\mu$ in Eq.~(\ref{15}) for the supertranslated Schwarzschild black hole by numerically solving the Eqs.~(\ref{7}), and then associating it with Eq.~(\ref{15}). It shows that the photon sphere, which locates at $r=3M$ (the black circle) for the bald black hole, is distorted (the red circle) for the supertranslated Schwarzschild black hole. The orange curves represent the light rays that are almost tangent to the photon sphere. One can observe that these light rays are determined by integrating constant $\kappa=27M^2$, which are the same as those in bald Schwarzschild space-time.  

\begin{figure*}[!h]
	\centering
	\includegraphics[scale=.54]{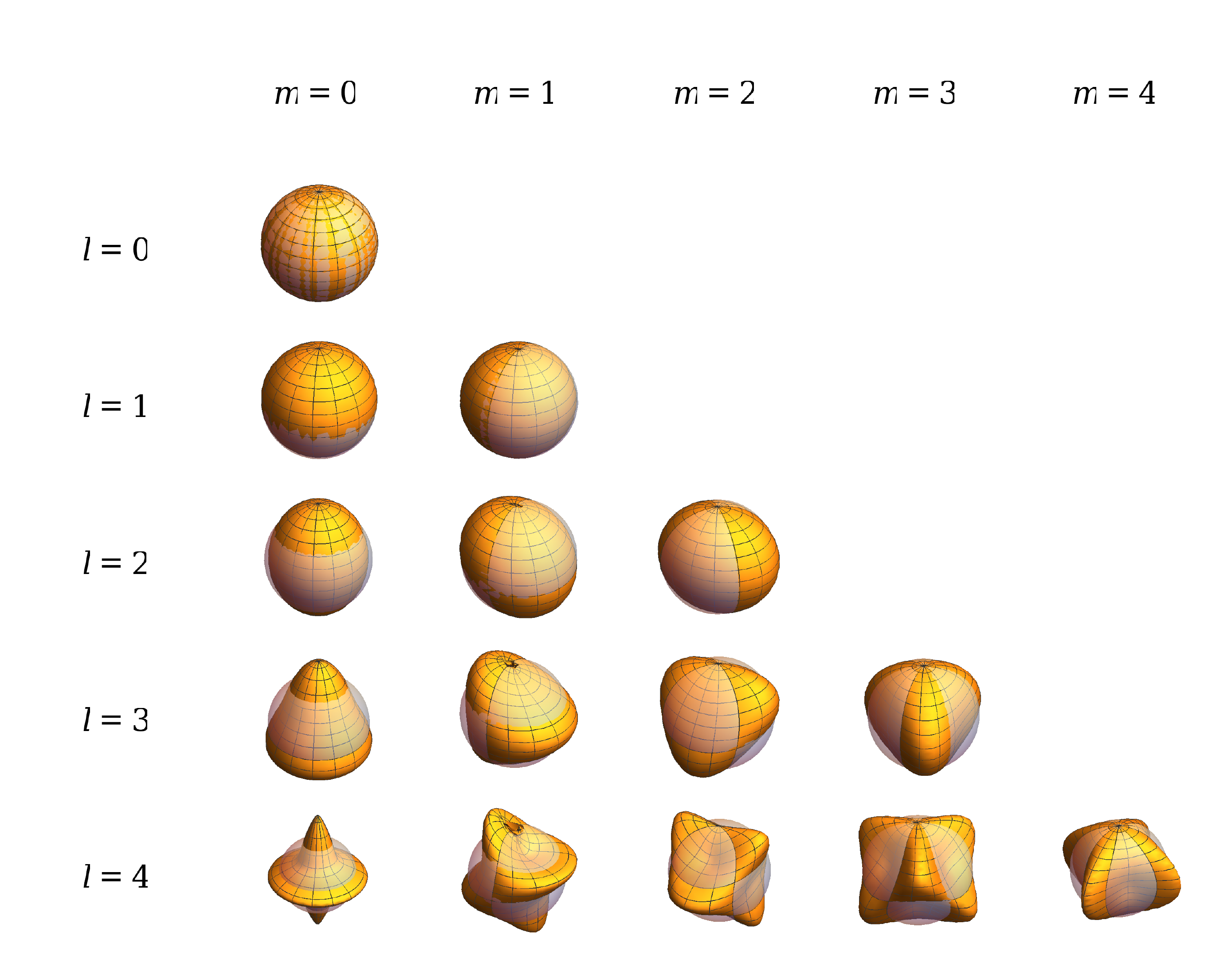}
	\caption{Distorted photon spheres of the supertranslated Schwarzschild black holes described by Eq.~(\ref{2}) and (\ref{4}) with $a_{lm}=\frac{1}{5}$. For comparison, the photon sphere of the bald Schwarzschild black hole is plotted in the translucent sphere. \label{F2}}
\end{figure*}

Based on Eq.~(\ref{15}), one can also obtain the distorted photon sphere for other types of supertranslation hairs. In Fig.~\ref{F2}, we plot the distorted photon spheres for selected $l$ and $m$ in Eq.~(\ref{4}). It shows that the photon sphere get more distorted for the larger $l$. Because the positive and negative $m$ in Eq.~\ref{4} share the same shape of the photon sphere, we only show the cases of $m \geq 0$ for illustration.

\smallskip
\section{The shadow of a supertranslated Schwarzschild black hole \label{IV}}
For the supertranslated Schwarzschild black hole, it was suggested that the distorted photon sphere might lead to a distinguishable shadow \cite{Sarkar:2021djs}.
In this section, we will calculate the shadow  in details. For distant observers, the shadow described by the coordinates $(\alpha,\beta)$ on the observers' projection plane is given by \cite{Synge:1966okc,Bardeen:1973tla} 
\m
\alpha&\equiv& - \frac{r\sin\theta \rd \phi}{\rd r}\bigg|_{x^i_{\rm obs}} = -r_{o}\sin\theta_{o}\frac{p^{{\phi}, (\rm hairy)}_{c}}{p^{{r}, (\rm hairy)}_{c}}\bigg|_{x^i_{\rm obs}}\nonumber\\&=&-\frac{\lambda +2 \partial_{\phi}f(\theta_{o},\phi_{o})}{r_{o}\sin\theta_{o}}~,\label{12}\\
\beta&\equiv& \frac{r \rd \theta}{\rd r}\bigg|_{x^i_{\rm obs}} = r_{o} \frac{p^{{\theta}, (\rm hairy)}_{c}}{p^{{r}, (\rm hairy)}_{c}}\bigg|_{x^i_{\rm obs}}\nonumber\\&=&\frac{\sqrt{\kappa-\lambda^2 \csc^2 \theta_{o}}+2  \partial_{\theta}f(\theta_{o},\phi_{o})}{r_{o}}~,\label{13}
\n
where $x^i_{\rm obs}=(r_o,\theta_o,\phi_o)$ is the location of an observer, and we have used Eq.~(\ref{12-1}) for evaluating the 4-velocities $p^{\mu, (\rm hairy)}_c$. Here, the $\kappa$ and $\lambda$ are determined by the critical escape orbits, and the values of them are exactly the same as those for bald Schwarzschild space-time as discussed in Eq.~(\ref{12-1}). 

\begin{figure}[!h]
	\includegraphics[scale=0.65]{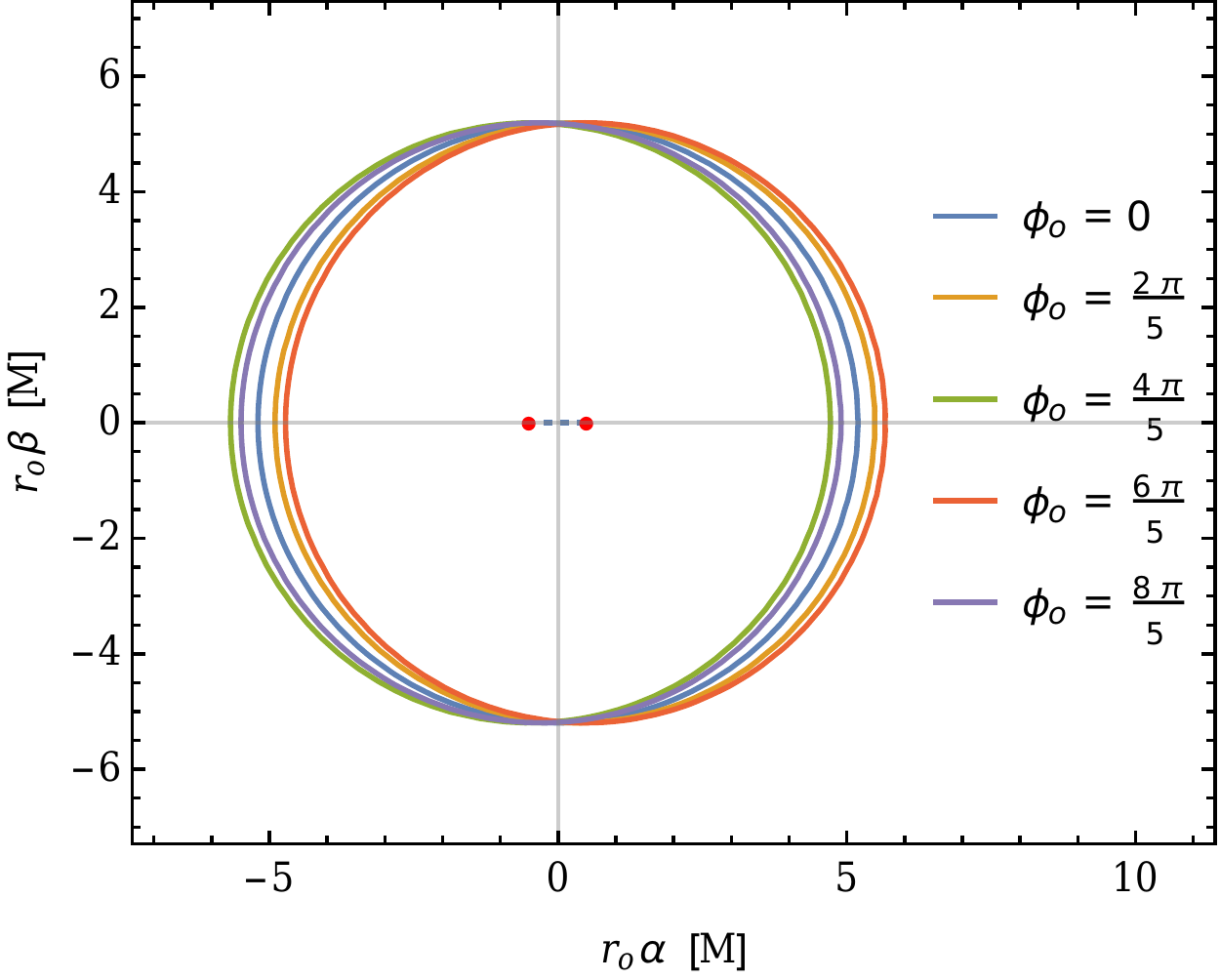}
	\caption{The shadow of supertranslated Schwarzschild black hole with $f(\theta,\phi)=\frac{1}{5}Y_{33}(\theta,\phi)$ for distant observers at equatorial plane $\theta_0=\frac{\pi}{2}$, and at different $\phi_o$ ranged from $0$ to $2\pi$, respectively. The photon rings on the projection plane are presented in the black circles, and red point--dashed line--red point represents the the centre of the shadows. \label{F4}}
\end{figure}

\begin{figure}[!h]
	\includegraphics[scale=0.65]{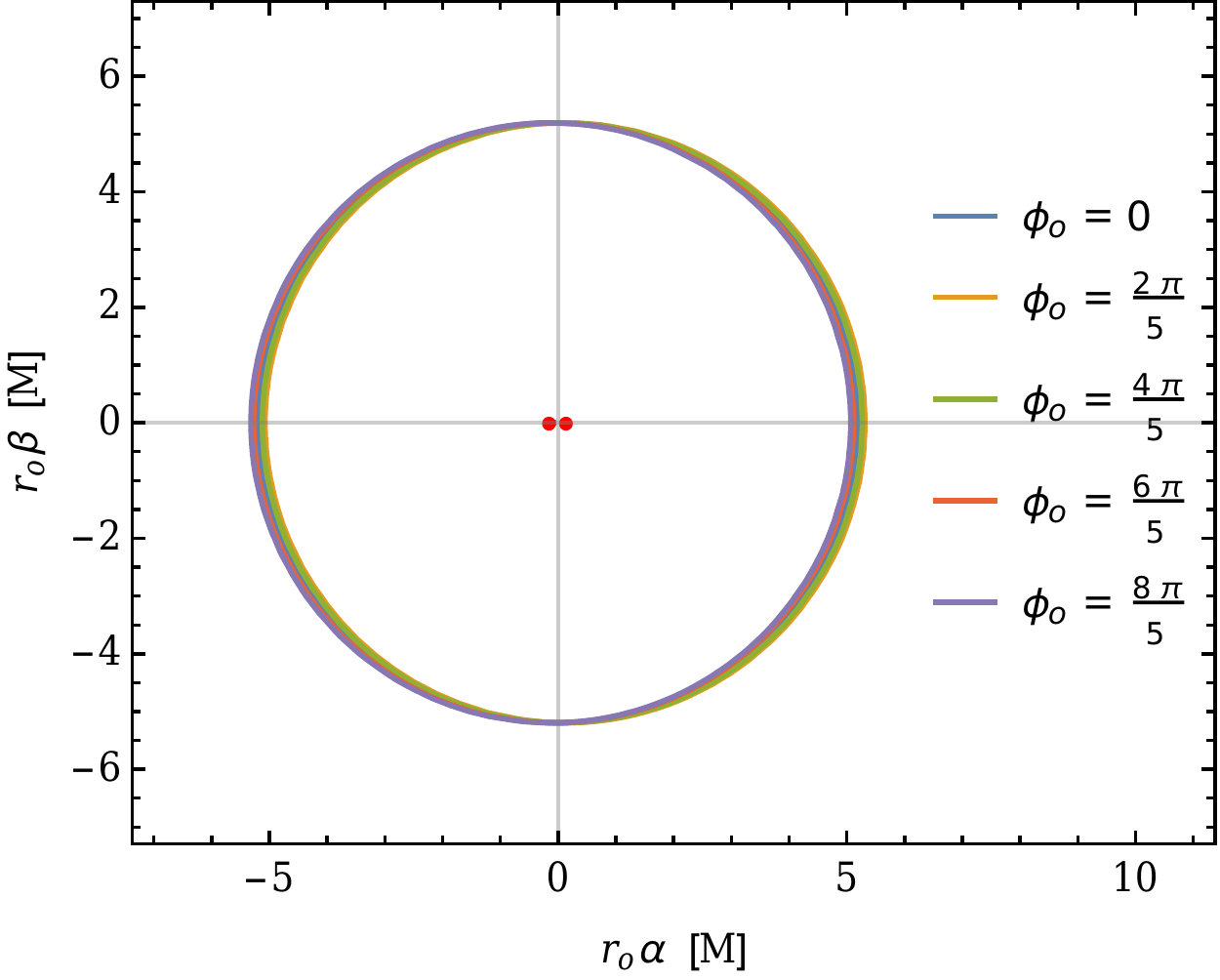}
	\caption{The shadow of supertranslated Schwarzschild black hole with $f(\theta,\phi)=\frac{1}{5}Y_{31}(\theta,\phi)$ for distant observers at equatorial plane $\theta_0=\frac{\pi}{2}$, and at different $\phi_o$ ranged from $0$ to $2\pi$, respectively. The photon rings on the projection plane are presented in the black circles, and red point--dashed line--red point represents the the centre of the shadows.   \label{F3}}
\end{figure}

\begin{figure}[!h]
	\includegraphics[scale=0.65]{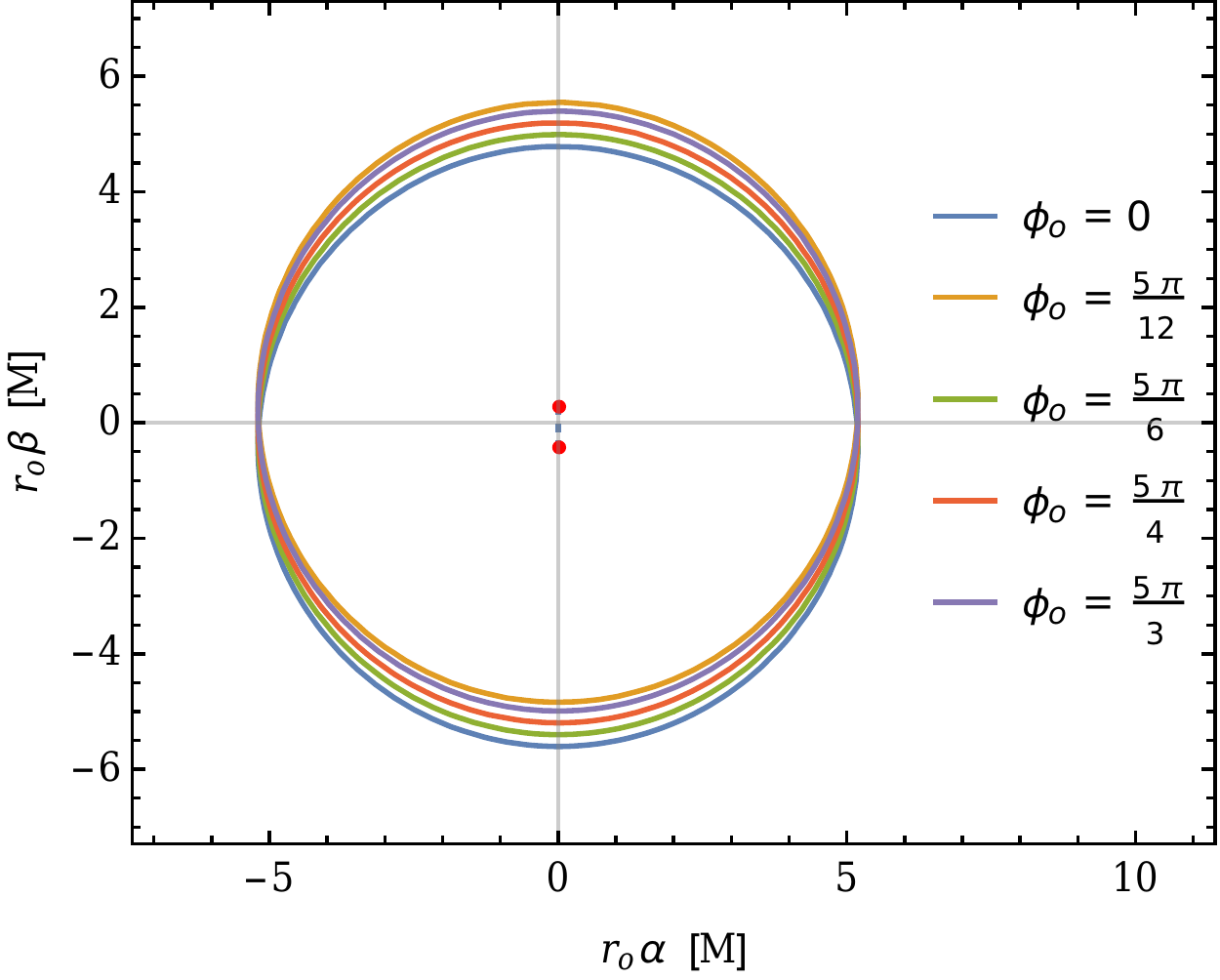}
	\caption{The shadow of supertranslated Schwarzschild black hole with $f(\theta,\phi)=\frac{1}{5}Y_{32}(\theta,\phi)$ for distant observers at equatorial plane $\theta_0=\frac{\pi}{2}$, and at different $\phi_o$ ranged from $0$ to $2\pi$, respectively. The photon rings on the projection plane are presented in the black circles, and red point--dashed line--red point represents the the centre of the shadows.   \label{F5}}
\end{figure}

Based on Eqs.~(\ref{12}) and (\ref{13}), the shape and size of the shadow remain the same as those for bald Schwarzschild black hole, but the position of the shadow is shifted by $\delta\alpha=-2\partial_\phi f/(r_0\sin\theta_0)$ and $\delta\beta=2\partial_\theta f/r_0$ on the projection plane. It can be clearly shown by eliminating the $\lambda$ in Eqs.~(\ref{12}) and (\ref{13}), and then one can obtain the curve equation in coordinates $(\alpha,\beta)$, 
\m
\left(\alpha+\frac{2 \partial_{\phi}f}{r_{o}\sin\theta_{o}}\right)^2+\left(\beta-\frac{2  \partial_{\theta}f}{r_{o}}\right)^2=\frac{\kappa}{r^2_{o}}~.\label{14}
\n
The curve equation represents a circle with radius $\sqrt{\kappa}/r_0$ and centre $(-2\partial_\phi f/(r_0\sin\theta_0),2\partial_\theta f/r_0)$ on projection plane. For illustrations, we plot the shadows of the supertranslated Schwarzschild black hole with $f(\theta,\phi)=\frac{1}{5}Y_{33}(\theta,\phi)$ and $f(\theta,\phi)=\frac{1}{5}Y_{31}(\theta,\phi)$ for observers at equatorial plane $\theta_0 = \frac{\pi}{2}$ in Figs.~\ref{F4} and \ref{F3}, respectively. For observers located at different $\phi$ ranged from $0$ to $2\pi$, the center positions of the shadows would move from the red point to the other one shown in the plots. Comparing the two examples, one might find that the positions of the shadows with supertranslation hair $f(\theta,\phi)=\frac{1}{5}Y_{33}(\theta,\phi)$ is shifted by a larger distance than that with $f(\theta,\phi)=\frac{1}{5}Y_{13}(\theta,\phi)$. In Fig.~\ref{F5}, we plot the shadows of the supertranslated Schwarzschild black hole with $f(\theta,\phi)=\frac{1}{5}Y_{32}(\theta,\phi)$ for observers at equatorial plane. Differed from above two cases, the observer move around the supertranslated Schwarzschild black hole would view the shadows shifted along the orbital axis. In addition, the positions of the shadows for $f(\theta,\phi)=\frac{1}{5}Y_{30}(\theta,\phi)$ are fixed at the origin of the coordinate $(\alpha,\beta)$ on projection plane. Thus, we did not show the case here. Generally, for arbitrary types of supertranslation hairs described in Eqs.~(\ref{2}) and (\ref{4}), the shadows on projection plane are completed described by Eq.~(\ref{14}). Here, we only showed the supertranslated Schwarzschild black hole with $l=3$ for examples.

\smallskip
\section{Conclusions and Discussions \label{V}}

We investigate the shadow for supertranslated Schwarzschild black hole  proposed by Hawking, Perry and Strominger \cite{Hawking:2016msc}. The calculation about the shadow of supertranslated Schwarzschild black hole turns to be much easier, if the supertranslated 4-velocities and the trajectories of the light rays are adopted. Based on this approach, we showed that the photon sphere gets distorted due to the supertranslation hairs, and the position of the shadow on the projection plane is shifted by the supertranslation vector and depends on the location of the observer, but the size and shape of the shadow are the same as those of  Schwarzschild black hole.



In this paper, we mainly focus on the shadow for distant observers. It might be non-trivial to consider the observer located at finite distance \cite{Grenzebach:2014fha,Stuchlik:2018qyz,Chang:2020miq,Tsupko:2019pzg,Li:2020drn}, in particular for those near the event horizon \cite{Donnay:2018ckb}. 
In this case, the distorted photon sphere might also have impact on the shape or size of the shadows.




\smallskip
{\it Acknowledgments. } 
This work is supported by the National Key Research and Development Program of China Grant No.2020YFC2201502, grants from NSFC (grant No. 11975019, 11991052, 12047503), Key Research Program of Frontier Sciences, CAS, Grant NO. ZDBS-LY-7009, CAS Project for Young Scientists in Basic Research YSBR-006, the Key Research Program of the Chinese Academy of Sciences (Grant NO. XDPB15).

\bibliography{ref}
	
\end{document}